\documentclass[acmsmall]{acmart}

\usepackage{algorithmic}
\usepackage{graphicx}
\usepackage{enumitem}
\usepackage{textcomp}
\usepackage{listings}
\usepackage{makecell}
\usepackage{longtable}
\usepackage{float}
\usepackage{graphicx}
\usepackage{multirow}
\usepackage{balance}
\usepackage{hyperref}
\usepackage{tabularx} 
\usepackage{booktabs} % For formal tables
\usepackage{colortbl}
\usepackage{hhline}
\usepackage{geometry}
\usepackage[dvipsnames,svgnames,x11names]{xcolor}
\usepackage[markup=underlined, commandnameprefix=always]{changes}
\usepackage[most]{tcolorbox} % For the fancy box
\usepackage{wrapfig}         % For wrapping text around figures/boxes
\usepackage[most]{tcolorbox}
\usepackage{todonotes}
\usepackage{pifont}         % for ding symbols
\usepackage{tikz}
\usetikzlibrary{arrows.meta, shapes.geometric, fit, positioning}

\usepackage[utf8]{inputenc}

 \pdfoutput=1 

\setcopyright{cc}
\setcctype{by-nc-nd}
\acmDOI{10.1145/3808130}
\acmYear{2026}
\acmJournal{PACMSE}
\acmVolume{3}
\acmNumber{FSE}
\acmArticle{FSE123}
\acmMonth{7}
\acmSubmissionID{fse26mainb-p349-p}
\received{2025-09-10}
\received[accepted]{2026-03-24}

\reversemarginpar
\setcommentmarkup{\todo[color={authorcolor!20},size=\scriptsize]{#3: #1}}

\newcommand{\cmark}{\textcolor{green}{\ding{51}}}
\newcommand{\xmark}{\textcolor{black}{\ding{55}}}

\newcommand{\subsc}[1]{\noindent \textbf{#1}}

\definechangesauthor[color=BrickRed]{SB}
\definechangesauthor[color=OliveGreen]{JS}
\definechangesauthor[color=RoyalPurple]{AA}

\begin{abstract}
 Toxic interactions during code reviews can undermine teamwork and hinder productivity in software engineering (SE) teams. While prior studies explore toxicity detection and empirical investigation, they lack real-time detoxification tools to support the SE community. To address this gap, we present ToxiShield, a browser extension for GitHub pull requests that is built using three modules: i) Toxicity Filter -- to identify whether a text is toxic, ii) Communication coach -- to facilitate just-in-time fine-grained toxicity categorization with explanations, and iii) The Reframer -- that generates a revised, constructive alternative of a toxic text. For each module, we trained and evaluated multiple deep learning and Large Language Models (LLMs) to identify the best choice. A BERT-based binary detection model, trained on 38,761 code review samples, achieves $98\%$ accuracy and an F1-score of $97\%$ and is the selected one for the Toxicity Filter \revision{module}. For the Communication Coach, prompt-tuned Claude 3.5 Sonnet achieved the best performance with $39\% MCC$ and $42\% F1$  in multiclass toxicity classification with detailed reasoning. For Reframer, we evaluated five LLMs using a fine-tuning strategy on a dataset of 10,120 code review comments. The fine-tuned Llama 3.2 model achieves $95.27\%$ style transfer accuracy, $97.03\%$ fluency, $67.07\%$ content preservation, and an $84\%$ J-score. We further validated ToxiShield through a human evaluation using the Technology Acceptance Model with 10 participants, confirming its perceived usefulness and ease of adoption. ToxiShield sets a benchmark for advancing constructive communication in software engineering, driving inclusivity and healthier collaboration in open-source communities.
 
\end{abstract}

\newcolumntype{L}[1]{>{\raggedright\let\newline\\\arraybackslash\hspace{0pt}}m{#1}}
\newcolumntype{C}[1]{>{\centering\let\newline\\\arraybackslash\hspace{0pt}}m{#1}}
\newcolumntype{R}[1]{>{\raggedleft\let\newline\\\arraybackslash\hspace{0pt}}m{#1}}
\usepackage{subfigure}
\usepackage{enumitem}

\newenvironment{boxedtext}
    {
    
    \begin{center}

    \begin{tabular}{|p{0.96\linewidth}|}
    \hline
    }
    { 
    \\ \hline
    \end{tabular} 
    
    \end{center}
       }

\newcommand{\revision}[1]{\textcolor{black}{#1}}

\newenvironment{revisionenv}
  {\begingroup\color{black}}
  {\endgroup}

\begin{document}

\title{ToxiShield: Promoting Inclusive Developer Communication through Real-Time Toxicity Filtering}

\author{Md Awsaf Alam Anindya}
\orcid{0009-0002-0866-3696}
\affiliation{%
  \institution{Bangladesh University of Engineering and Technology}
  \city{}
  \country{Bangladesh}
}
\email{1505114.maaa@ugrad.cse.buet.ac.bd}

\author{Showvik Biswas}
\orcid{0009-0006-2154-4498}
\affiliation{%
  \institution{Bangladesh University of Engineering and Technology}
  \city{}
  \country{Bangladesh}
}
\email{1805068@ugrad.cse.buet.ac.bd}

\author{Anindya Iqbal}
\orcid{0000-0002-2763-8819}
\affiliation{%
  \institution{Bangladesh University of Engineering and Technology}
  \city{}
  \country{Bangladesh}
}
\email{anindya@cse.buet.ac.bd}

\author{Jaydeb Sarker}
\orcid{0000-0001-6440-7596}
\affiliation{%
  \institution{University of Nebraska Omaha}
  \city{}
  \country{USA}
}
\email{jsarker@unomaha.edu}

\author{Amiangshu Bosu}
\orcid{0000-0002-3178-6232}
\affiliation{%
  \institution{Wayne State University}
  \city{}
  \country{USA}
}
\email{amiangshu.bosu@wayne.edu}

\begin{abstract}
 Toxic interactions during code reviews can undermine teamwork and hinder productivity in software engineering (SE) teams. While prior studies explore toxicity detection and empirical investigation, they lack real-time detoxification tools to support the SE community. To address this gap, we present ToxiShield, a browser extension for GitHub pull requests that is built using three modules: i) Toxicity Filter -- to identify whether a text is toxic, ii) Communication coach -- to facilitate just-in-time fine-grained toxicity categorization with explanations, and iii) The Reframer -- that generates a revised, constructive alternative of a toxic text. For each module, we trained and evaluated multiple deep learning and Large Language Models (LLMs) to identify the best choice. A BERT-based binary detection model, trained on 38,761 code review samples, achieves $98\%$ accuracy and an F1-score of $97\%$ and is the selected one for the Toxicity Filter \revision{module}. For the Communication Coach, prompt-tuned Claude 3.5 Sonnet achieved the best performance with $39\% MCC$ and $42\% F1$  in multiclass toxicity classification with detailed reasoning. For Reframer, we evaluated five LLMs using a fine-tuning strategy on a dataset of 10,120 code review comments. The fine-tuned Llama 3.2 model achieves $95.27\%$ style transfer accuracy, $97.03\%$ fluency, $67.07\%$ content preservation, and an $84\%$ J-score. We further validated ToxiShield through a human evaluation using the Technology Acceptance Model with 10 participants, confirming its perceived usefulness and ease of adoption. ToxiShield sets a benchmark for advancing constructive communication in software engineering, driving inclusivity and healthier collaboration in open-source communities.
 
\end{abstract}


%\ccsdesc[500]{Collaboration in software development~Toxicity}
%\ccsdesc[500]{Software and its engineering~Open source model}
%\ccsdesc[500]{Software and its engineering~Programming teams}

%\keywords{toxicity, software engineering, large language model}

\renewcommand{\shortauthors}{Anindya \em{et} al.}


\maketitle

\section{Introduction}
\label{sec:introduction}

Successful Open Source Software (OSS) projects require strong coordination and teamwork among contributors~\cite{sowe2008understanding, trinkenreich2021pots}. Developers primarily collaborate using text-based channels, including mailing lists, code reviews, version control systems, and bug trackers. Despite this need for collaboration, research indicates that communication within OSS communities frequently becomes unprofessional and toxic~\cite{raman2020stress, sarker2020benchmark, miller2022did, sarker2023automated, ehsani2023exploring, sarker2025landscape, imran2026toxicity}.

This hostility produces severe consequences. It creates significant barriers for new participants~\cite{raman2020stress} and causes active contributors to leave the project~\cite{sarker2020benchmark}. Furthermore, toxic interactions disproportionately affect minority and junior developers~\cite{gunawardena2022destructive}. Such behavior leads to emotional distress and burnout, which subsequently diminish developer productivity~\cite{raman2020stress, rahmandwhp}. Over time, these interpersonal conflicts degrade the overall health of the community by undermining diversity~\cite{miller2022did, gunawardena2022destructive} and sustaining ongoing disputes~\cite{egelman2020predicting, rahmandwhp}. Consequently, this toxic environment ultimately compromises the quality of the software produced~\cite{miller2022did}.

To counter these issues, many OSS projects have adopted Codes of Conduct (CoC) to encourage respectful environments; however, their practical impact remains limited~\cite{murphy2024did,bharadwajshifting}. The primary obstacle is enforcement. Current moderation typically relies on manual validation, a labor-intensive process that is unsustainable for large-scale communities~\cite{miller2022did}. The inadequacy of manual moderation has spurred the development of automated systems designed to identify and mitigate harmful discourse. Consequently, a growing body of Software Engineering (SE) research focuses on creating tools to detect not only overt toxicity~\cite{raman2020stress, sarker2023automated, sarker2023toxispanse} but also nuanced antisocial behaviors, including incivility~\cite{rahmandwhp,ferreira2024incivility}, destructive criticism~\cite{gunawardena2022destructive}, sexism~\cite{sultana-sgid-2025,dev-esem-2025}, and pushback~\cite{egelman2020predicting}.

Nevertheless, existing approaches for managing toxic interactions in OSS are primarily reactive. They operate on a post hoc basis, using classifiers to flag problematic text for later review and removal~\cite{sarker2023automated}. This delay is critical because, by the time an intervention occurs, significant damage to contributor relationships may have already been done~\cite{GitHubOpenSourceSurvey2017,rahmandwhp}. To foster genuinely healthier communication, a paradigm shift is necessary from reactive flagging to proactive prevention. Inspired by real-time grammar checkers in platforms like Grammarly, we aim to develop a tool integrated directly into developers' communication channels. This tool would not only flag potentially toxic text during composition but also provide actionable feedback, explaining why the text is problematic and suggesting neutral or civil alternatives. While recent work by Rahman~\textit{et al.} moved in this direction by developing a customized Text-to-Text Transfer Transformer (T5) model to rephrase uncivil comments, their solution has two major limitations. First, it lacks the real-time integration necessary for proactive moderation. Second, and more importantly, it fails to explain its reasoning. This explanatory capability is essential for fostering long-term behavioral change and helping users navigate the cultural nuances of toxic communication~\cite{kumar2021}.

\begin{figure}[t!]
	\centering  
\includegraphics[width=\linewidth, trim=0 0 0 0, clip] {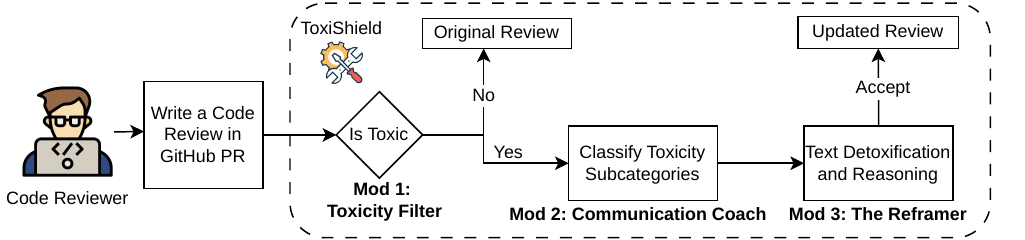}
	\caption{Motivational Workflow of ToxiShield}
	\label{fig:method_flow}	
    \vspace{-12pt}
\end{figure}

To address these limitations, we propose \textbf{ToxiShield}, a comprehensive framework designed to proactively detect and mitigate toxicity in SE communication, with a specific focus on code review interactions. Unlike post hoc solutions, ToxiShield operates in real-time as a developer drafts feedback. The system identifies and mitigates potential toxicity through a three-stage process, as illustrated in Figure~\ref{fig:method_flow}, comprising the following key modules.

\vspace{2pt} \noindent \textbf{Module 1: Toxicity Filter}: Acting as the first line of defense, this module performs a binary classification to distinguish between toxic' and non-toxic' code review comments. When a comment is flagged as toxic, the filter automatically triggers a downstream workflow, routing the problematic text to specialized modules for granular analysis and remediation.

\vspace{2pt} \noindent \textbf{Module 2: Communication Coach}: The Communication Coach functions as an automated pedagogical agent aimed at fostering long-term behavioral improvement. Upon receiving a comment identified as toxic, it conducts a detailed analysis to classify the text into specific subcategories of toxicity, such as Insult, Threat, or Obscenity. The module then generates a targeted explanation detailing precisely why the comment is problematic, thereby transforming a standard moderation event into a constructive learning opportunity for the user.

\vspace{2pt} \noindent \textbf{Module 3: The Reframer}: This module serves as a generative remediation assistant designed to detoxify harmful discourse. When a comment is flagged, The Reframer analyzes the author's core intent and generates a revised, constructive alternative that preserves the technical message while adhering to professional standards. This suggested revision is presented to the user alongside a brief rationale, explaining the specific changes made and why the new version is more effective for maintaining a collaborative environment.

To identify the optimal architecture for each module, we trained and evaluated multiple \revision{machine learning} models. Our evaluation reveals that for the primary detection task, lightweight models offer superior performance, surpassing even Large Language Models (LLMs) such as GPT-4o. Specifically, the \textit{BERT-base-uncased} model achieved the highest performance for binary toxic text detection, attaining an accuracy of \revision{98\%} and an $F1$-score of \revision{97\%}. For the more granular task of classifying toxicity subcategories,
\revision{Claude 3.5 Sonnet yielded the best overall results, achieving a macro $F1$-score of 42\% (average $F1$-score of 75\%) and a macro Matthews Correlation Coefficient (MCC) of 39\% (average MCC of 66\%)}.  Finally, regarding detoxification, our fine-tuned Llama 3.2 model outperformed other state-of-the-art LLMs, achieving a J-Score of 84\%. Beyond raw metrics, these techniques successfully generated high-quality rationales for the suggested changes, demonstrating the efficacy of language models in addressing complex linguistic challenges. We validated our approach through a comprehensive evaluation comprising both automated metrics and user studies to ensure the generated alternatives maintained the original context and intent while effectively reducing toxicity. A user study involving 10 software developers, evaluated via the Technology Acceptance Model (TAM), confirmed across four key metrics that our extension effectively aids in reducing toxicity in code reviews. The key contributions of this study are as follows:

\begin{itemize}[leftmargin=*]
\item A curated dataset of toxic code reviews paired with their corresponding non-toxic alternatives.
\item A robust multiclass toxicity classifier capable of categorizing specific types of toxicity.
\item The development of \textbf{ToxiShield}, a browser extension-based tool for code review detoxification. Our replication package, including code, data, and experimental results, is available at ~\cite{ToxiShield}. 

\end{itemize}

The remainder of this paper is organized as follows. 
Sections~\ref{sec:toxicity-detection}, \ref{sec:multiclass}, and \ref{sec:toxicity-detoxification} detail the design and evaluation of the three core modules of ToxiShield, respectively. 
Section~\ref{sec:browser-etenstion} presents the results of our integrated user evaluation. 
We discuss our findings in Section~\ref{sec:discussion}, address threats to validity in Section~\ref{sec:threats}, review related works in Section~\ref{sec:background}, and conclude the paper in Section~\ref{sec:conclusion}.

\section{Module 1: Toxicity Filter}
\label{sec:toxicity-detection}
\begin{figure}
    \centering
    \includegraphics[width=1\linewidth]{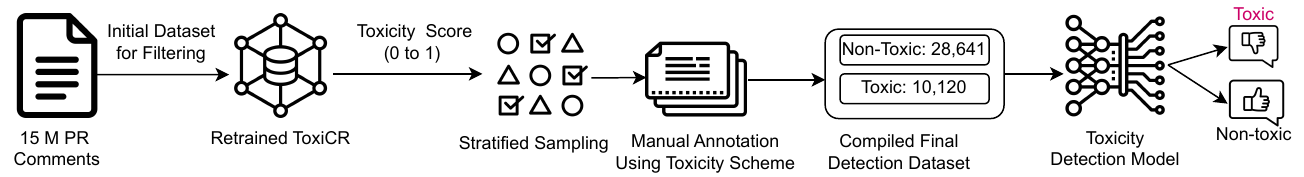}
    \caption{\revision{Workflow of the Toxicity Filter Module where 15 M PR comments are selected from ~\cite{sarker2025landscape}.}}
    \label{fig:detection-model}
    \vspace{-10pt} 
\end{figure}

A robust binary toxicity detector serves as the foundational entry point of our pipeline, acting as a critical filter for all incoming comments. Effective binary filtration is essential to prevent the downstream classification and detoxification modules from being inundated with non-toxic content, thereby optimizing computational efficiency and ensuring system responsiveness. While prior research has introduced binary toxicity detectors for the SE domain~\cite{sarker2023automated, raman2020stress, rahmandwhp}, these approaches often prioritize offline classification accuracy over the latency constraints required for seamless integration into real-time developer workflows. For instance, the current state-of-the-art tool, ToxiCR~\cite{sarker2023automated}, achieves strong performance metrics ($88.9\%$ F1-score, $95.8\%$ accuracy); however, recent studies indicate that its efficacy diminishes when processing nuanced or sarcastic expressions~\cite{rahmandwhp}, which are common in collaborative coding environments. Consequently, the primary objective of this module is to accurately classify code review comments as either toxic or non-toxic, addressing these limitations. Figure~\ref{fig:detection-model} illustrates the methodology adopted for Module 1, the design and evaluation of which are detailed in the following subsections.

\subsection{Dataset Construction}
\label{sec:dataset}

\subsc{\revision{Conceptualization of Toxicity in Code Reviews}}. 
\label{sec:conceptualization-toxicty} 
Recent research within the OSS community has documented a broad spectrum of antisocial behaviors, ranging from general toxicity~\cite{miller2022did, sarker2023automated} to specific manifestations such as destructive criticism~\cite{gunawardena2022destructive}, pushback~\cite{egelman2020predicting}, and incivility~\cite{ferreira2021shut}. In this work, we target \textit{toxicity} within code reviews. We adopt the operational definition of toxicity provided by Sarker \textit{et} al.~\cite{sarker2025landscape}, whose taxonomy synthesizes prior SE findings~\cite{miller2022did, ferreira2021shut, sarker2023automated} into 11 distinct categories. Therefore, within the context of this study, a text is considered toxic if it falls into one of the following categories: \textit{Profanity, Trolling, Insult, Self-deprecation, Entitlement, Identity Attack, Threats, Obscenity, Arrogance, Flirtation}, or \textit{Object-Directed Toxicity}. Consequently, behaviors such as \textit{sarcasm} and \textit{condescension} are not designated as toxic solely on the basis of their expression, as these behaviors may serve benign social functions or lack a standardized definition in SE taxonomies. We therefore label them as toxic in our binary dataset only when they meet the definitional criteria for one of the 11 core subcategories. For example, the comment \textit{``Grandma, what long lines you have?''} is classified as toxic not merely due to its sarcastic tone, but because it constitutes a veiled insult devoid of constructive feedback.

\vspace{2pt}
\subsc{Dataset Selection}
To the best of our knowledge, the largest existing binary toxicity dataset in the SE domain is the ToxiCR corpus~\cite{sarker2023automated}, which contains 19,651 code review comments. However, this dataset exhibits a significant class imbalance; it is heavily skewed towards profanity (comprising nearly half of the samples), leaving other forms of toxicity underrepresented. Furthermore, alternative datasets with toxicity labels are limited in scale~\cite{raman2020stress, ehsani2024incivility}. To construct a more diverse and representative dataset, we leveraged the corpus from a recent study by Sarker \textit{et} al.~\cite{sarker2025landscape}, which mined 101 million comments across 2,828 GitHub-based OSS projects. Aligning with our specific focus on code reviews, we isolated Pull Request (PR) comments from this repository. After excluding 86 million issue comments, we were left with approximately 15 million PR comments.

\vspace{2pt}
\subsc{Dataset Filtering} Given the sparsity of toxic comments in OSS discussions~\cite{sarker2020benchmark, Qiu2022}, automated classification is essential for efficient large-scale dataset curation. To improve detection capabilities, we retrained the ToxiCR model using a composite dataset merging the Sarker \textit{et} al. corpus (19,651 samples)~\cite{sarker2023automated} and the Rahman \textit{et} al. dataset (4,700 samples)~\cite{rahmandwhp}. This retraining process significantly enhanced the model's robustness and calibration regarding the nuances of code review discourse. We then deployed this retrained model across the 15 million PR comments, generating a probability score (0 to 1) for each entry. While the standard ToxiCR framework classifies text as toxic at a threshold of $\ge 0.5$~\cite{sarker2023automated}, prior work indicates this threshold yields a high false negative rate for underrepresented toxicity classes~\cite{rahmandwhp}. Consequently, we expanded our candidate search space to include probabilities ranging from 0.10 to 1.0. This broader range allows us to capture a more diverse spectrum of toxicity inherent to code reviews.

\vspace{2pt}
\subsc{Dataset Labeling} To mitigate potential labeling bias, two annotators were trained on the conceptual framework detailed in Section~\ref{sec:conceptualization-toxicty}. We employed a stratified sampling strategy, dividing the unlabeled data into five distinct bins based on the toxicity probability ($p$) assigned by the model (see Table~\ref{tab:dataset}). Because the model tends to assign high probabilities primarily to explicit toxicity (e.g., profanity), this stratification enabled us to systematically target implicit or subtle toxicity in the lower-probability bins. From each bin, the first annotator manually reviewed entries until 2,100 toxic samples were identified, resulting in an initial corpus of 10,500 candidates. A second annotator independently validated these selections, removing 380 instances deemed non-toxic. This rigorous validation process yielded a final, high-quality dataset of 10,120 toxic samples.

\begin{table}[t]
\centering
    \caption{\revision{Dataset Annotation with Examples:  10,120 toxic and 28,641 non toxic samples in the final dataset. }}
    %\revisionenv{ 
%\color{blue}
%{
    \begin{tabular}{|p{2.2cm}|p{6.5cm}|p{4cm}|}
\hline

\textbf{Probability} &\textbf{ Example of PR comments} & \textbf{Rationale} \\ \hline
0.10 $\le$ p < 0.28 & Please try to understand the code you're commenting on. & Represents Condescension (Implicit Toxicity) \\ \hline
    0.28 $\le$ p < 0.46 & Oh, it's called Trainwreck. Surprised you have you not heard of it? & Represents Sarcasm \\ \hline
    0.46 $\le$ p < 0.64 & you are a liar!? & Insulting comment \\ \hline
    0.64 $\le$ p < 0.82 & That's just wacky. As I said in the issue report, bizarre that it would even miscompile in debug mode in MSVC. & Expresses Object-Directed Toxicity (Dismissive Tone) \\ \hline
    0.82 $\le$ p $\le$ 1.0 & Yep, fuck those technomancers, always hated. & Expresses Profanity \\ \hline

\end{tabular}
%}

%}

 \begin{comment}

Probability Ranges & Examples & Rationale   \\ \hline
 $0.10 \le p < 0.28$ &Please try to understand the code you're commenting on.  & Represents Condescension \\ \hline
  $0.28 \le p < 0.46$ & Oh, it's called Trainwreck. Surprised you have you not heard of it? & Represents Sarcasm \\ \hline

$0.46 \le p < 0.64$ &you are a liar!? & Insulting comment \\ \hline

$0.64 \le p < 0.82$ &That's just wacky. As I said in the issue report, bizarre that it would even miscompile in debug mode in MSVC  & This comment expresses Object-Directed Toxicity. \\ \hline
$0.82 \le p < 1.0$ &Yep, fuck those technomancers, always hated them too. & Expresses Profanity \\ \hline

% Something tells me his sister is Beelzebub.

\end{comment}
    \label{tab:dataset}
\end{table}

\vspace{2pt}
\subsc{Non-toxic Samples}
To compile the non-toxic component of the dataset, we randomly selected 30,000 samples from the pool of 15 million PR comments, which were subsequently reviewed and confirmed as non-toxic by the first annotator. To further ensure data purity, we applied a keyword-based filtering process to these candidates, removing 1,359 samples containing explicit profanity. This rigorous validation yielded a final set of 28,641 non-toxic samples. Combining these with our toxic corpus resulted in a comprehensive dataset of 38,761 samples, comprising \textbf{10,120 toxic} and \textbf{28,641 non-toxic} PR reviews. Crucially, by selecting samples across the full spectrum of ToxiCR's~\cite{sarker2023automated} probability distribution, we mitigate the risk of inheriting the model's existing biases in our subsequent evaluation.

%\vspace{10pt}
\subsection{Model Training} 
\label{detection-ft}

To identify the optimal architecture for ToxiShield's Toxicity Filter, we evaluated a diverse set of candidate models, detailed in Table~\ref{tab:detection}, using the dataset described in Section~\ref{sec:dataset}. Our selection strategy encompassed both Masked Language Models (MLMs) and Large Language Models (LLMs). Specifically, we fine-tuned \textit{BERT-base-uncased}~\cite{bert} and \textit{Xtreme-DistilBERT}~\cite{xtremedistilbert}, and incorporated the pre-trained ToxiCR model~\cite{sarker2023automated} to serve as a domain-specific baseline. Additionally, we employed GPT-4o and GPT-3.5 as LLM-based detectors, following the benchmarking methodology of Rahman~\textit{et al.}~\cite{rahmandwhp}.
For the experimental setup, the dataset was partitioned into training (80\%), validation (10\%), and testing (10\%) sets. We initialized the MLMs with their publicly available pre-trained weights and appended a task-specific classification head to each. Input sequences were tokenized with a maximum length of 128 tokens. Fine-tuning was conducted over 12 epochs using a learning rate of $2 \times 10^{-5}$, a batch size of 128, and a weight decay of 0.01, with the objective of minimizing binary cross-entropy loss. To ensure robustness, we applied 10-fold stratified cross-validation within the training split and monitored per-epoch validation performance to select the checkpoint that maximized the toxic-class F1-score for final testing.

\subsection{Model Evaluation}

\begin{table*}[t]
    
    \caption{Performance of binary toxicity detection models. Bold indicates the highest score per metric.}
    
\label{tab:detection}
    \centering
    
\begin{tabular}{|p{5.7cm}|r|r|r|r|r|r|r|}
\hline
\small
 
\multirow{2}{*}{{\textbf{Models}}} & 

\multicolumn{3}{c|}{{\textbf{Non-toxic}}} & \multicolumn{3}{c|}{{\textbf{Toxic}}} & \multirow{2}{*}{{\textbf{Accuracy}}} \\ \cline{2-7}

 &   {$P_0$} & {$R_0$} & $F1_0$ & {$P_1$} & {$R_1$} & $F1_1$  & \\ \hline

  \textbf{BERT-base-uncased (fine-tuned)}          & \textbf{0.98} & \textbf{0.99} & \textbf{0.99} & \textbf{0.98} & \textbf{0.96} & \textbf{0.97} & \textbf{0.98}       \\ \hline
 Microsoft/Xtreme DistilBert (fine-tuned) & \textbf{0.98}          & 0.98          & 0.98          & 0.97          & 0.93          & 0.95                                      & \textbf{0.98}   \\ \hline
 
 GPT-4o (prompted) & 0.84 & 0.99 & 0.91 & 0.96 & 0.48 & 0.64 & 0.86  \\ \hline

 ToxiCR (off-the-shelf)                     & 0.95          & 0.89          & 0.92         & 0.74          & 0.87          & 0.80        & 0.87                       \\ \hline

\end{tabular}
    \vspace{-10pt}
\end{table*}

To evaluate our detection models, we employed standard classification metrics: precision, recall, accuracy, and the F1-score. Given the critical necessity of accurately identifying toxic content in code reviews, we designated the F1-score of the toxic class as our primary performance metric. Table~\ref{tab:detection} presents a comparative performance analysis of our fine-tuned models, \textit{BERT-base-uncased} and \textit{Xtreme-DistilBERT}, against established baselines, including the off-the-shelf ToxiCR model and a zero-shot prompted GPT-4o. Among the evaluated architectures, the fine-tuned \textit{BERT-base-uncased} model demonstrated superior performance, achieving the highest precision (0.98) and recall (0.96), with a toxic-class F1-score of 0.97. \textit{Xtreme-DistilBERT} followed closely with an F1-score of 0.95. These results underscore the robustness of \textit{BERT-base-uncased} in identifying toxic discourse. Consequently, we selected it as the default detection engine for ToxiShield's Toxicity Filter module.

\subsection{Error Analysis of Toxicity Filter Module}

We conducted an error analysis on our best model, \textit{BERT-base-uncased}, which misclassified 59 of 3,877 test samples (23 False Positive (FPs), 36  False Negatives (FNs)). Using an open-coding approach based on established SE research~\cite{sarker2023automated, sultana-sgid-2025}, two authors categorized these failures into four types.  

\vspace{2pt}
\noindent {\textbf{i) Pragmatic and Contextual Misinterpretation:} 
The model often failed to detect sarcasm, passive-aggressiveness, and mockery, accounting for most misclassifications (65\% of FPs, 81\% of FNs). Positive words frequently masked toxic intent, causing FNs. For instance, the model prioritized ``hilarious'' over the mocking tone in: ``This means you can make pizza-flavoured ice cream, you tell me that isn't a hilarious notion''. Conversely, it detected FPs by over-indexing on negative words during valid technical criticism, incorrectly flagging: ``this thing is still just terrible, average of 9s''.

\vspace{2pt}
\noindent \revision{\textbf{ii) Domain-Specific Terminology Misinterpretation:} These errors arise when standard technical identifiers, jargon, or references are misidentified as general English profanity or aggression. This category accounted for $26\%$ of FPs and $11\%$ of FNs. A PR comment \textit{``convert anal.autoname into `int'}'', which was incorrectly flagged as toxic (FP). The model failed to recognize \textit{``anal}'' as a standard abbreviation for ``\textit{analysis}'' within the software domain context.}

\vspace{2pt}
\noindent \textbf{iii) Misinterpretation of Self-Deprecation:} 
The model occasionally conflated benign self-deprecation, a common, non-toxic behavior in code reviews~\cite{sarker2025landscape, sarker2023automated, miller2022did, kadiyala2026should}, with hostility (9\% of FPs, 5\% of FNs). For instance, it incorrectly flagged ``i dig it. will do when I'm not drunk'' (FP), likely reacting to the keyword ``drunk'' rather than the humor.

\vspace{2pt}
\noindent \revision{\textbf{iv) Obfuscation and Tokenization Failures:} We identified a single instance ($3\%$ of FNs) where the model failed due to adversarial text modification designed to evade filters. The comment ``\textit{Yes there is, i ju- F U C K.}'' was classified as non-toxic (FN). In this case, the insertion of spaces between characters obfuscated the profanity, causing a tokenization failure that prevented accurate detection.}

\vspace{4pt}
 \begin{boxedtext}
\textbf{Key Finding 1:} 
\textit{Fine-tuned \textit{BERT-base-uncased} demonstrated the best performance in detecting toxic and non-toxic comments for code review text, when trained on our new dataset.}
\end{boxedtext}

\section{Module 2: Communication Coach}
\label{sec:multiclass}
The Communication Coach is an automated pedagogical module designed to address the nuances of toxic behavior in software engineering. Because toxicity encompasses diverse forms like insults and entitlement~\cite{miller2022did, sarker2025landscape}, binary classification often oversimplifies the problem and obscures specific harms that disproportionately affect underrepresented groups~\cite{ehsani2025analyzing}.  Moving beyond simple flagging upon detection (Section~\ref{sec:toxicity-detection}), Communication Coach classifies the comment into a subcategory and provides targeted explanatory feedback. This granular approach transforms standard moderation into a constructive learning opportunity, fostering transparency and long-term behavioral change. The following sections detail the design and evaluation of this multi-class classification module.

\subsection{Dataset for Toxicity Subcategories}

Prior SE studies have identified various subcategories of toxicity~\cite{miller2022did, ehsani2024incivility, sarker2025landscape}. Among these, Sarker~\textit{et al.}~\cite{sarker2025landscape} offer the most comprehensive taxonomy, identifying 11 distinct toxicity types in GitHub pull requests. Given the scarcity of fine-grained annotated data, we utilized their dataset of 600 samples, which contains 532 verified toxic instances. To create a balanced distribution for inference, we supplemented this core with 600 non-toxic comments drawn from our baseline dataset (Section~\ref{sec:toxicity-detection}). The resulting multiclass dataset comprises 1,200 code review comments ({532 toxic samples and 668 non-toxic samples}) spanning 12 distinct categories, including the non-toxic class. Among the 532 toxic samples, 140 instances are labeled with two distinct subcategories (e.g., \textit{Profanity} and \textit{Insult}), and three instances are labeled with three subcategories, consistent with the primary labeling schema of Sarker \textit{et} al.~\cite{sarker2025landscape}. We explicitly preserve these overlapping annotations, thereby formulating the problem as a multi-label, multi-class classification task~\cite{dekel2010multiclass}.

\subsection{Prompt-based Design For LLMs}

To classify toxicity subcategories, we leveraged LLMs, which have demonstrated strong performance in prompt-based multi-class tasks~\cite{sun2023text, wangzeroshotclass2023, Gretz2023}. While our detection module relies on MLMs for computational efficiency (Section~\ref{sec:toxicity-detection}), we transitioned to LLMs for this phase to exploit their generative reasoning capabilities. We evaluated state-of-the-art models from the OpenAI, Claude, and Llama families, employing iterative prompt optimization to maximize classification accuracy~\cite{sun2023text, wangzeroshotclass2023, Guo2024, Roy2023}. 

A pivotal design decision in this phase was the adoption of \textit{In-Context Learning} (prompting) over fine-tuning models like Llama 3.2. This choice was driven by the severe data scarcity inherent to fine-grained toxicity classification. Robust fine-tuning necessitates a substantial volume of diverse examples to ensure generalization. However, our subcategory dataset contains only 532 toxic instances distributed across 11 distinct categories, with long-tail classes such as \textit{Arrogance} and \textit{Obscenity} containing negligible samples. Such a limited volume is insufficient for fine-tuning multi-billion parameter models without inducing severe overfitting~\cite{touvron2023llama, Hu2021-kn}. Consequently, we leveraged prompt engineering, a strategy that has demonstrated efficacy in both general toxic text classification~\cite{Guo2024, Roy2023} and recent SE-specific multiclass tasks~\cite{dev-esem-2025}. This approach enables us to harness the sophisticated reasoning capabilities of LLMs while circumventing the high data requirements of parameter optimization.

\noindent \textbf{Choice of LLMs:} 
We conducted a comprehensive evaluation of LLMs in our dataset. Our model selection includes three families of models, covering both proprietary and open-source systems: GPT4o-mini, GPT 4o, Claude 3.5 Sonnet, Claude 3.7 Sonnet, Llama 3.1 70B, and Llama 3.3 70B. 
%All detailed results are available in our replication package~\cite{ToxiShield}.

\vspace{2pt}
\subsc{Prompt Schema and Components:} \label{sec: prompt-schema}
To operationalize the subcategory classification approach, we designed prompts capable of classifying comments into single or multiple toxicity categories or identifying them as non-toxic. The structure of our baseline prompt is defined as follows:

    \begin{tcolorbox}[
        colback=teal!5!white,    % Light teal background
        colframe=teal!75!black,  % Dark teal frame border
        title=\textbf{Elements of Basic Prompt for Multiclass Classification},
        arc=2mm                  % Rounded corners
    ]
    
    \begin{enumerate}[leftmargin=*,labelindent=0pt]
        \item \textbf{Role Clarification:} 
        You are a maintainer of an OSS project. Your task is to classify the subcategories of toxicity. 
                
        \item \textbf{Task Assignment:} Please assign one or more toxicity categories to the input comment and to provide a brief rationale (step-by-step).
        
        \item \textbf{Contextualized Definitions \& Few-Shot Examples:} To mitigate semantic drift, we provided concise, iteratively refined definitions for all 11 subcategories (e.g., \textit{Identity Attack}, \textit{Entitlement}). Each definition was paired with at least one domain-specific example to illustrate the boundary between technical feedback and toxicity.

        \item \textbf{Classification Guidelines:} If the comment is not toxic, respond with ``Non-Toxic'' and a supporting explanation.
        
        \item \textbf{Output Format Specification:} Respond in XML format: \verb|<response> {response} </response> <category> {category} </category>|. Now classify the following comments. 
    \end{enumerate}
    \end{tcolorbox}

    \label{fig:multiclass-prompt} 

This prompt is structured into five strategic components designed to optimize LLM performance: (i) role clarification, (ii) task assignment, (iii) category definitions supplemented by representative examples, (iv) explicit operational guidelines, and (v) a strict XML output specification. We designate this initial configuration as \textit{Prompt 1}, which serves as the baseline for subsequent iterative refinements. Consistent with our binary classification framework (Section~\ref{sec:toxicity-detection}), we align our subcategory definitions directly with the taxonomy established by Sarker \textit{et} al.~\cite{sarker2025landscape}. Comprehensive details regarding the prompt architecture and its evolution are provided in our replication package~\cite{ToxiShield}.

\vspace{2pt}
\subsc{Prompt Optimization and Iterative Refinement}
Prompt engineering has emerged as a critical discipline for maximizing LLM performance in domain-specific tasks~\cite{Saravia_Prompt_Engineering_Guide_2022}, significantly reducing the cognitive load on developers~\cite{khurana2024and}. Within ToxiShield, we implemented a unified prompting framework across two key modules: (i) The Communication Coach (Section~\ref{sec:multiclass}), which assigns specific toxicity subcategories, and (ii) The Reframer (Section~\ref{sec:toxicity-detoxification}), which rewrites toxic comments into professional alternatives. Our approach is grounded in three core principles: structured reasoning,  iterative validation against quantitative metrics, and schema-constrained outputs.

First, to enhance the model's inferential capabilities~\cite{takeshi_kojima_2022_llm_reasoning}, we adopted Chain-of-Thought (CoT) prompting~\cite{wei2022chain}. This technique decomposes the complex tasks of classification and detoxification into discrete, logical steps. Specifically for classification, CoT compels the model to articulate the rationale behind a specific label, thereby improving both accuracy and interpretability. Second, to mitigate hallucinations and reduce variability in toxicity subcategory classification, we employed a systematic, iterative prompt optimization strategy. Aligning with established methodologies for prompt-based multi-class classification~\cite{sun2023text, wangzeroshotclass2023, Guo2024, Roy2023, dev-esem-2025}, we refined our prompts through five distinct evolutionary stages (Figure~\ref{fig:multiclass-prompt-evolution}). This evolution progressed from generic instructions to precise, persona-anchored directives. We explicitly conditioned the model to adopt the persona of an experienced OSS maintainer, ensuring it could distinguish constructive technical criticism from actual toxicity. At each stage, performance was rigorously assessed against a validation set using the Matthews Correlation Coefficient (MCC), a metric widely recognized for its robustness in multi-class classification tasks~\cite{dev-esem-2025}. We specifically selected GPT-4o for the prompt optimization phase because its advanced instruction-following capabilities make it an ideal engine for structuring complex classification tasks. The five stages in our prompt optimization are as follows.

\begin{itemize}[leftmargin=*]

\item \textit{Stage 1 -- The Baseline (Zero-Shot):} We commenced with a foundational zero-shot prompt that provided only class names and high-level classification instructions. This stage established a performance baseline, evaluating the model's inherent capabilities prior to the introduction of domain-specific guidance (Section~\ref{sec: prompt-schema}).

\item \textit{Stage 2 -- Operational Definitions (Observable Cues):} Manual analysis revealed that the model struggled with abstract definitions. Consequently, we re-framed the prompt to prioritize observable behavioral cues. Key iterations included: (i) replacing abstract definitions with actionable, behavior-based criteria for each subcategory; (ii) integrating canonical few-shot examples to ground the model's understanding of complex classes; (iii) explicitly distinguishing overlapping categories (e.g., \textit{Insult} vs. \textit{Trolling}) to resolve edge cases; and (iv) imposing strict output constraints to ensure parsing reliability on the full test set.

\item \textit{Stage 3 -- Detecting Subtle Nuance:} To address the model's difficulty in detecting implicit toxicity, we refined the prompt to capture subtle nuances such as condescension and sarcasm. Iterations involved: (i) explicitly defining \textit{Sarcasm} and \textit{Irony} to prevent the literal interpretation of positive words in negative contexts; (ii) introducing initial lexical markers for strong sentiment; and (iii) expanding constraints to account for variations in offensive slang and non-standard profanity.\item 

\textit{Stage 4 -- Lexical and Rule-Based Constraints:} While Stage 3 improved tone detection, the model still missed toxicity driven by specific vocabulary. We addressed this by integrating lexical resources and logic rules: (i) embedding a curated list of profanity terms as high-probability toxicity indicators; (ii) incorporating an Anger List (e.g., capitalization patterns, aggressive punctuation) to detect heightened emotional states; (iii) applying negative constraints to reduce false positives arising from technical disagreements; and (iv) implementing critical logic rules (e.g., \texttt{IF profanity is used positively, DO NOT label as Insult}) to enforce consistency.

\item \textit{Stage 5 -- Rare Category Refinement:} Evaluations of Stage 4 indicated persistent under-performance in rare categories, specifically \textit{Arrogance} ($N=5$) and \textit{Identity-Attack} ($N=17$). We performed targeted refinement by expanding these definitions with granular cues and additional representative examples to improve recall.

\end{itemize}

\begin{figure}
    \centering
    \includegraphics[width=\linewidth]{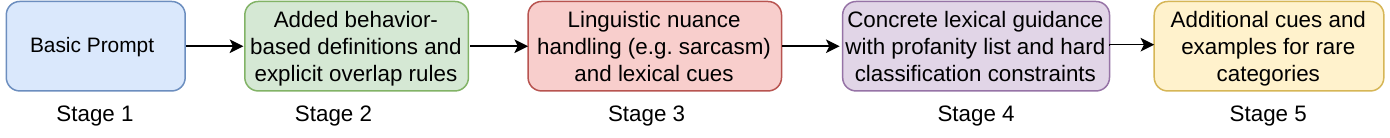}
    \caption{Communication Coach: Iterative Prompt Evolution}
    \label{fig:multiclass-prompt-evolution}
    \vspace{-20pt}
\end{figure}

We selected the final prompt based on validation performance while preserving the basic prompt components. Table~\ref{tab:prompt-refinement-multiclass} details the iterative evolution of our prompt design with the GPT-4o model, highlighting the performance bottlenecks encountered, the specific refinements applied, and the resulting trajectory of the model's performance. From the prompt evolution, we observed that Stage 4 achieved the best macro MCC score, even though it missed one rare category, ``Arrogance''. We have used this best prompt for further validation of other models. 

\begin{table*}[t]
\centering
    \caption{Prompt Refinement with Macro MCC scores for Communication Coach}
    \begin{revisionenv} 
    \begin{tabular}{|p{0.5cm}|p{4.5cm}|p{6.5cm}|r|}
\hline

\textbf{Stg.} & \textbf{Example} & \textbf{Misclassification Reason} & \textbf{MCC} \\ \hline
1 & Let’s just remove this for now ... kill all the commented-out source. & Prompt has only abstract definitions and ``kill" is misclassified as a threat. & 0.31 \\ \hline
2 & Helluva test. I had to look at it for a while before it oozed through. Nice job! & Prompt does not contain explicit instructions to handle sarcasm. & 0.30 \\ \hline
3 & @Imagebard feck off & Prompt does not identify ``feck" as a profane word due to a misspelling, and it not having a list of commonly misspelled profanities. & 0.31 \\ \hline
4 & Go ahead and merge this nonsense. I can work around it. & Prompt does not have cues in it that can help the model detect subtle hints of arrogance. & 0.37 \\ \hline

5 & Oh boy this is sexy :smiley\_cat: & Prompt misinterprets comment as obscene due to the presence of ``sexy", a common marker for sexual content, and its definition of an obscene comment having any sexual content or references to sexual acts. & 0.35 \\ \hline

\end{tabular}

\end{revisionenv}
    \label{tab:prompt-refinement-multiclass}
\end{table*}

\subsection{Evaluation of Communication Coach}

\textbf{Evaluation Metric} 
We measured multiclass model performance using exact match (EM), precision, recall, F1, and MCC scores \revision{under two aggregations: (i) \textbf{Avg} (aggregate over all instances) and (ii) \textbf{Macro} (unweighted average across 12 caegories)}, as presented in Table~\ref{table:multi-class}.

\begin{table*}[t]
    \caption{Comparison of LLMs for Multi‑Class Classification For Best Model in Each Family}
\label{table:multi-class}
    \centering
    \begin{tabular}{|l|c|c|c|c|c|c|c|c|c|}
\hline
 &  & \multicolumn{2}{c|}{\textbf{Precision}} & \multicolumn{2}{c|}{\textbf{Recall}} & \multicolumn{2}{c|}{\textbf{F1}} & \multicolumn{2}{c|}{\textbf{MCC}} \\ \cline{3-10}
\textbf{Model Name} & \textbf{EM}
& Avg & Macro 
& Avg & Macro 
& Avg & Macro 
& Avg & Macro \\ 
\hline

GPT 4o 
& \textbf{0.69} 
& {0.72} & 0.41 
& \textbf{0.77} & \textbf{0.42}
& {0.74} & 0.40
& {0.65} & 0.37
\\ \hline

Claude 3.5 Sonnet 
& 0.63 
& \textbf{0.74} & 0.44
& 0.75 & \textbf{0.42}
& \textbf{0.75} & \textbf{0.42}
& \textbf{0.66} & \textbf{0.39} \\ \hline

Llama 3.3 70B 
& 0.61 
& 0.69 & 0.43 
& 0.74 & 0.36
& 0.71 & 0.37 
& 0.60 & 0.34 \\ \hline

\end{tabular}

\end{table*}

\noindent \textbf{Evaluation Results} 
Table~\ref{table:multi-class} presents the multi-class classification performance of the leading models from each LLM family (GPT-4o, Claude 3.5 Sonnet, and Llama 3.3 70B). For this final evaluation, we utilized the optimal prompt (Stage 4) identified in Table~\ref{tab:prompt-refinement-multiclass} and employed deterministic decoding (temperature = 0) to ensure reproducibility. Overall, Claude 3.5 Sonnet demonstrated the most balanced detection capability across the 12 categories, achieving the highest Macro F1-score (0.42) alongside a Macro MCC of 0.39. GPT-4o performed competitively (Macro F1: 0.40, MCC: 0.40), while Llama 3.3 70B exhibited lower efficacy (Macro F1: 0.37, MCC: 0.34). Interestingly, while Claude provided superior balance across minority classes, GPT-4o favored exact match accuracy 0.69. These findings highlight considerable variability in how contemporary LLMs handle imbalanced, multi-label toxicity categorization.

\subsection{Error Analysis of Communication Coach}

\subsc{Confusion Matrix}
To assess the detection performance of the best model (Claude 3.5 Sonnet) with the most optimized prompt on multiclass classification, we conducted an evaluation in Figure~\ref{fig:multiclass-heatmap}. To rigorously resolve the ambiguity in the multi-label ground truth for this analysis, a human rater manually assigned the single most appropriate label to each sample (i.e., total sample size: 1200) from the best model's output. The results reveal a sharp contrast: while explicit categories like Profanity (168/245) and Non-Toxic (605/668) proved highly robust, the model struggled with ambiguous and rare classes. A primary failure mode occurred in Object-Directed Toxicity (OD Toxicity), which was frequently misclassified as Non-Toxic, Profanity, Insult, or Trolling, indicating difficulty distinguishing technical frustration from actual toxicity. Furthermore, extreme data sparsity in rare interpersonal traits such as Arrogance ($N=2$) and Entitlement ($N=17$) severely hindered detection, underscoring a critical need for dataset expansion and refined category definitions.

\begin{figure}
    \centering
    \includegraphics[width=1\linewidth]{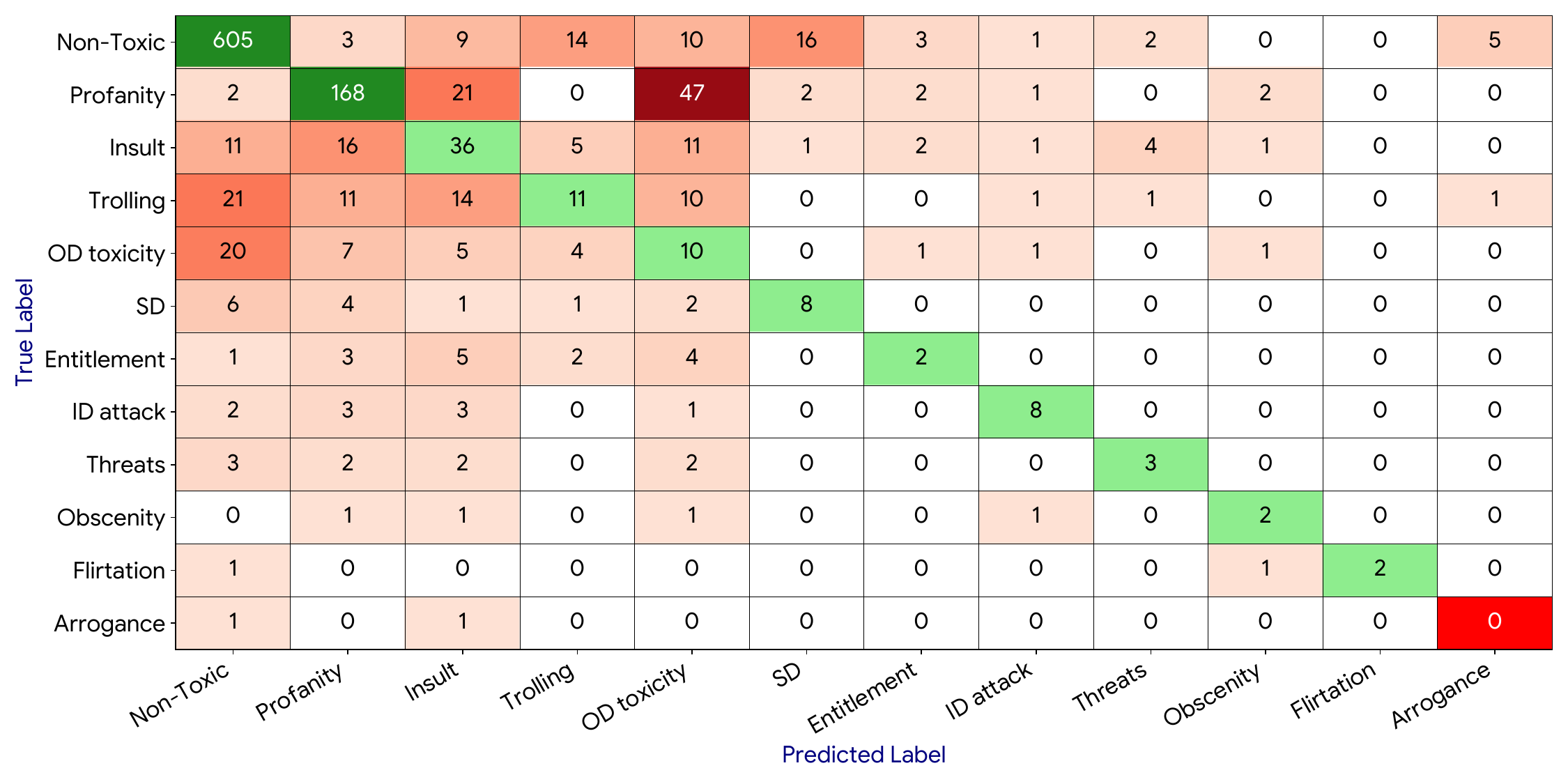}
    \caption{Confusion Matrix for Communication Coach, where SD represents Self Deprecation, Identity Attack represents ID Attack. Values in the diagonal (greener shades) are correct classifications.}
    \label{fig:multiclass-heatmap}
    \vspace{-10pt}
\end{figure}

\vspace{2pt}
\subsc{Manual Evaluation:}
To complement our quantitative metrics and real-world performance, we conducted a manual evaluation of our best-performing multiclass model, Claude 3.5 Sonnet. Two authors independently reviewed a random sample of 100 instances (44 non toxic and 56 toxic samples) drawn from the full dataset of 1,200, categorizing each prediction into one of three outcomes: correct, where the model correctly identified a comment as non-toxic or its exact subcategory, alternate, where the either partially predicted or misclassified the ground-truth labels, and incorrect, where a model misclassified a toxic comment as non-toxic or vice versa. Inter-rater reliability, measured via Cohen’s $\kappa$, was $0.66$, indicating substantial agreement~\cite{Landis1977}. Of the 100 evaluated instances (44 are non-toxic and 56 are toxic), $58$ were correctly classified, $25$ were strictly misclassified, and $17$ were assigned an alternate toxic label. Our annotators also evaluated the model's generated explanations for its classifications. The model performed particularly well at identifying and justifying overt toxicity. For instance, when correctly classifying \textit{``and that FUCKIN BRANCH NAME}'' as Profanity, the model accurately explained it as \textit{``Explicit use of profanity ("FUCKIN") to emphasize a point about a branch name''.}  However, the model struggled with nuanced or overlapping subcategories. For example, \textit{``Frigging love you Niki. Seriously}'' was misclassified as non-toxic despite its original Flirtation label. Interestingly, the model's explanation accurately recognized the text as ``Highly informal and affectionate language'' yet it still applied a non-toxic label. Similarly, \textit{``how dare u accuse me of trying to freeload off other ppl}'' was flagged as Insult and Trolling, though it was originally labeled as Threat. These instances demonstrate that while the model reliably detects general toxicity and generates logical rationales, it occasionally struggles to resolve the blurred boundaries between closely related interpersonal toxicities.

\begin{boxedtext}
\textbf{Key Finding 2:} 
\textit{ Claude 3.5 Sonnet achieves best performance on multiclass toxicity classification with $0.39$ MCC and $0.42$ F1 score. However, for nuanced or data-sparse toxicity types, models fail to capture subcategories efficiently, which requires a more balanced dataset.}
\end{boxedtext}

\section{Module 3: The Reframer}
\label{sec:toxicity-detoxification}

\begin{figure}
    \centering    \includegraphics[width=.9\linewidth]{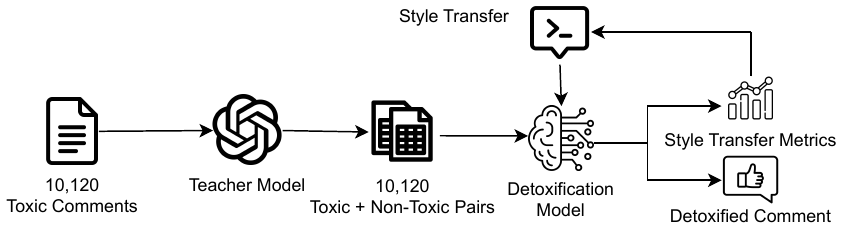}
    \caption{\revision{The Reframer: A teacher model is used to create a parallel dataset, which is used to train a student detoxification model with iterative prompting.}}
    \label{fig:placeholder2}

    \vspace{-10pt}
\end{figure}

While flagging toxicity is a prerequisite for mitigation~\cite{bharadwajshifting, sarker2025landscape}, interventions that focus solely on detection risk stigmatize contributors without offering a path to improvement. In collaborative developer workflows, it is far more effective to guide users toward constructive alternatives than to simply reject their contributions. However, achieving effective real-time detoxification is non-trivial; it requires a precise balance between neutralizing hostility and preserving semantic fidelity, ensuring that the technical substance of the feedback remains intact. To address this challenge, we developed a specialized detoxification module that automatically rephrases toxic communication in OSS environments.

\subsection{Evaluating Detoxification}
\label{sub:evalmetrics}
Following the methodology established by Ostheimer \textit{et al.} ~\cite{ostheimer-etal-2024-text}, we employed four complementary metrics to evaluate the efficacy of our detoxification models in capturing the critical trade-off among neutralizing toxicity, preserving technical intent, and maintaining linguistic fluency.

 \noindent \textbf{Style Transfer Accuracy (DETOX):} This metric quantifies the model's success in stripping toxicity from the text. We utilized ToxiCR~\cite{sarker2023automated} to compute toxicity probability scores for both the original and detoxified samples, calculating the net percentage reduction. As the primary objective of our tool is mitigation, high style accuracy serves as the critical indicator of success.

 \noindent\textbf{Fluency (FL):} We evaluated the grammatical correctness and naturalness of the detoxified output using a pre-trained CoLA (Corpus of Linguistic Acceptability) classifier~\cite{Warstadt2019-td}. High fluency scores indicate that the rephrased comments are syntactically sound and readable, a prerequisite for their acceptance in professional developer workflows.

 \noindent \textbf{Content Preservation (PRESERVE):} To assess semantic fidelity, we utilized the embedding-based similarity metric proposed by Wieting \textit{et} al.~\cite{wieting-etal-2019-beyond}. By encoding both the original and detoxified comments into paraphrastic sentence embeddings and computing their cosine similarity, this method captures conceptual equivalence rather than mere word overlap. Unlike rigid lexical metrics such as BLEU~\cite{papineni2002bleu}, this approach effectively credits valid paraphrases, ensuring that the technical meaning of the code review remains intact after detoxification.

 \noindent \textbf{J-Score:} To provide a holistic performance benchmark, we calculated J-Score, defined as the harmonic mean of Content Preservation, Fluency, and Style Accuracy. J-Score ensures high performance in one area (e.g., detoxification) is not achieved at the expense of others (e.g., semantic loss).

\subsection{Dataset Construction via Knowledge Distillation}
\label{sec:dataset_construction}

We formulate the detoxification of code reviews as a Text Style Transfer (TST) task, specifically aiming to shift the stylistic attribute from toxic to `neutral' or `professional' while rigorously preserving technical semantics. Effective TST requires a parallel corpus that pairs each toxic comment with a faithful, non-toxic rewrite, enabling models to learn content preservation alongside toxicity reduction~\cite{fu2018style, prabhumoye2018}. However, such a corpus does not exist for the software engineering domain. To address this scarcity, we employed a teacher-student knowledge distillation framework~\cite{hinton2015distilling} to synthesize a high-quality parallel dataset.

\vspace{2pt}
\subsc{Teacher-Generated Parallel Data:}
We utilized the 10,120 toxic comments identified in our filtering phase (Table \ref{tab:dataset}) as the source data. To generate their non-toxic counterparts, we leveraged multiple state-of-the-art LLMs acting as ``teacher'' models. These teachers were prompted to rewrite each toxic comment into a respectful alternative, strictly adhering to the original technical intent. This process yielded several candidate parallel datasets, one per teacher model.
To select the optimal training basis for our downstream (student) models, we evaluated each teacher-generated dataset using the quantitative metrics defined in Section~\ref{sub:evalmetrics}. The teacher model that maximized the aggregate J-Score was selected to construct the final parallel corpus.

\vspace{2pt}
\subsc{Prompt Engineering for Data Synthesis:}
The quality of the parallel dataset is directly dependent on the efficacy of the teacher's prompt. We iteratively refined these templates against our evaluation metrics until achieving stable performance. Detailed prompts are provided in our replication package~\cite{ToxiShield}. We optimized our prompts using several key strategies to ensure consistency and semantic fidelity:

\begin{itemize}[leftmargin=*]
\item \textit{Chain-of-Thought (CoT) and Few-Shot Learning:} To mitigate hallucination and improve reasoning, we employed CoT prompting combined with few-shot examples. This guided the teacher models to first analyze the toxic elements and then systematically reconstruct the sentence, ensuring the technical core remained intact.

\item \textit{Structured Output Constraints:} To facilitate automated parsing and integration, we enforced a strict output schema. The teacher models were instructed to generate responses in a standardized format, such as: ``\texttt{Detoxified: <rewritten comment>; Rationale: <explanation of changes>}''. This structure minimized output variability and ensured every rewritten comment was accompanied by an explanation, a critical feature of ToxiShield's pedagogical approach.
\end{itemize}

\vspace{2pt}
\subsc{Teacher Model Evaluation:}
To assess the quality of datasets generated by the candidate teacher models, we sampled 500 detoxified pairs from each dataset. We evaluated them using the metrics introduced in Section~\ref{sub:evalmetrics}. For each dataset, we report the average scores across these metrics along with the overall J-score, which summarizes their trade-offs. Table~\ref{tab:parallel_dataset} shows the evaluation metrics of the four candidate teacher models used in our study. While all teacher models demonstrated strong detoxification capabilities, their performance varied in striking a balance between meaning preservation and fluency. Among them, \textit{GPT 4o-05-13} achieved the highest J-score of $88.14\%$, indicating the best overall balance across metrics. We therefore selected the dataset generated by this model as the final parallel corpus for training the student models.

% \begin{table*}[!htpb]
%     \centering
%     \resizebox{\textwidth}{!}{%
%     \begin{tabular}{|c|c|c|c|c|l|l|} \hline 
%          Model& Toxicity Decrease (\%) &  Semantic Similarity (\%)&  Style Accuracy (\%)& Fluency (\%)&Cont Pre (\%) &J\\ \hline 
%          OpenAI GPT 3.5 Turbo&  91.74&  87.56&  99.80&  98.81&71.01 &88.43\\ \hline 
%          OpenAI GPT 4o-05-13 &  96.17&  85.37&  99.80&  99.01&73.86 &\textbf{89.65}\\ \hline 
%  Claude 3& 97.50& 79.65& 99.88& 98.73&57.41 &82.82\\ \hline 
%  Llama 3 -70B& 96.90& 84.54& 99.39& 95.08&65.51 &86.15\\ \hline
%     \end{tabular}
%     }
%     \caption{Translation metrics for different datasets generated by models.}
%     \label{tab:parallel_dataset}
% \end{table*}

\begin{table*}[!htpb]
\small
    \centering
     \caption{Translation metrics for teacher models used to generate parallel datasets}
    %\resizebox{\textwidth}{!}{%
    \begin{tabular}{|p{4.3 cm}|r|r|r|r|} \hline 
         \textbf{Model}& \textbf{DETOX (\%)}  & \textbf{FL (\%)} &\textbf{PRESERVE (\%)} &\textbf{J-Score (\%)} \\ \hline 
         OpenAI GPT 3.5 Turbo&  91.74& 98.81&71.01 &85.46\\ \hline 
         OpenAI GPT 4o-05-13 &  96.17& \textbf{99.01}& \textbf{73.86} &\textbf{88.14}\\ \hline 
 Claude 3& \textbf{97.50}& 98.73&57.41 &79.36\\ \hline 
 Llama 3 - 70B & 96.90& 95.08&65.51 &83.10\\ \hline
    \end{tabular}
    %}
   
    \label{tab:parallel_dataset}
\end{table*}

\subsection{Detoxification Model Training}
Following the identification of \textit{GPT-4o-2024-05-13} as the superior teacher model, we utilized its generated parallel corpus to fine-tune a lightweight, open-source student LLM. This student model serves as the core inference engine for ToxiShield's detoxification module. To optimize computational efficiency without compromising model performance, we employed Low-Rank Adaptation (LoRA)~\cite{Hu2021-kn}. \revision{The fine-tuning process consisted of 10 epochs using a Stratified 8:2 train: test split of the 10,120 samples, employing 10 epochs with a learning rate of $2 \times 10^{-4}$, a weight decay of $0.01$, and 5 warm-up steps}. All training procedures were executed on a cloud instance equipped with a single 24 GB NVIDIA RTX 4090 GPU. To ensure reproducibility and minimize stochasticity, we enforced deterministic generation parameters across all experiments (temperature $= 0.0$, maximum output length $= 256$ tokens). This strict control allows us to attribute performance variations directly to model architecture and prompt design rather than random sampling noise.

\subsection{Detoxification Model Evaluation}
We evaluated the performance of our distilled student models using the identical metric set applied in the teacher phase, ensuring direct comparability. Table~\ref{tab:student_metrics} summarizes the results across seven distinct LLM architectures. While the student models generally yielded lower raw scores than their teacher counterparts, several demonstrated exceptional trade-offs between toxicity reduction and semantic preservation. Most notably, \textit{Llama 3.2 3B} emerged as the top-performing candidate, achieving the highest aggregate J-Score of $84.00\%$. This model effectively balanced strong style transfer accuracy ($95.27\%$) and fluency ($97.03\%$) with respectable content preservation ($67.07\%$). \textit{Qwen 2.5 Instruct 7B} followed closely with a J-Score of $83.73\%$, while the proprietary \textit{GPT-4o mini} achieved a comparable $83.57\%$. In terms of specific component metrics, \textit{Gemma 2 2B} achieved the highest content preservation ($68.47\%$) but suffered from lower style transfer accuracy ($87.59\%$), whereas \textit{Phi 3.5} excelled in toxicity reduction ($94.16\%$) but lagged in content preservation ($64.77\%$). \revision{Given that the J-Score, the harmonic mean of these metrics, is the standard benchmark for TST tasks, these results confirm \textit{Llama 3.2 3B} as the optimal choice. Crucially, this finding highlights that a specialized, small-parameter open-source model can outperform larger proprietary models like \textit{GPT-4o mini} when fine-tuned for domain-specific detoxification.}

\begin{table}[t]
\small
    \caption{Performance comparison of detoxification models}
    \label{tab:student_metrics}
    \centering
    %\resizebox{\textwidth}{!}{
\begin{tabular}{|p{4.3 cm}|r|r|r|r|p{2 cm}|r|}

\hline

\textbf{Model}  & \textbf{DETOX (\%)} & \textbf{FL (\%)} & \textbf{PRESERVE (\%)} & \textbf{J-Score (\%)} \\ \hline
 
GPT-4o mini (off-the-shelf) & \textbf{96.35} & 98.36 & 65.14 & 83.57 \\  \hline
Llama 3.1 8B & 91.95 & 98.03 & 66.42 & 83.03 \\  \hline
% Llama 3.2 3B & Yes & 91.71 & 97.62 & 67.11 & 83.22 \\  \hline
Phi 3.5  & 94.16 & 96.48 & 64.77 & 82.36 \\ \hline
Gemma 2 2B  & 87.59 & 94.49 & 68.47 & 81.96 \\  \hline
Qwen 2.5 Instruct 7B  & 94.88 & 98.66 & 65.98 & 83.73 \\ \hline
GPT-3.5 FT (~\cite{rahmandwhp} off-the-shelf) & 88.50 & \textbf{99.31} & 59.38 & 78.51 \\ \hline
Llama 3.2 3B & 95.27 & 97.03 & \textbf{67.07} & \textbf{84.00} \\  \hline

\end{tabular}
\vspace{-12pt}
%}
\end{table}

\vspace{2pt}
\subsc{Error Analysis of Best Performing Model:}
To validate our model performance, we manually validate the misclassification of our model.
Specifically, we focus on where our model failed to rephrase the toxic comments. We labeled a detoxified comment as misclassified if it is scored 0.5 or above by ToxiCR. The best performing model, Llama 3.2 3B produced 78 misclassifications out of 2018 detoxified comments in the test set. We classify these instances into four categories.

\begin{itemize}[leftmargin=*]
    \item \emph{Name Error:} This type of error occurred when ToxiCR flagged otherwise benign comments as toxic due to the presence of profane substrings within usernames. Specifically, six such instances were observed, accounting for $7.69\%$ of the total misclassifications. For example, the comment ``@warunalakshitha, do we need to escape the dots?" was incorrectly labeled as toxic because the username contained a substring that matched a profane word.
\item \emph{Context Error:} The most prevalent category of errors is Context error, with 29 occurrences. Comments often had toxic text as names of variables, files, or folders. Our model did not rephrase them, as rephrasing them would change the context entirely. For example, \textit{``I'd change it to `is\_disgusting\_for`, as the current name implies that the item itself is disgusted by someone picking it up."} was identified as toxic due to the variable name containing the word \textit{disgusting}. 
\item \emph{General Error:} General errors are classified as those instances where our model failed to remove toxicity. \revision{For instance, \textit{``Resonators won't be shitty pickaxes now? That's great!"} where our model failed to detoxify it. There are 20 samples of this category in our sample.} 

\item \emph{ToxiCR Error:} These were solely due to the Toxicity Filter module misclassifying a detoxified comment as toxic. There were 23 such cases. For example, \textit{`::' is C++ code. Another example is `Device.enumerat-ed\_found'?} was identified as toxic by ToxiCR.

\end{itemize}

\vspace{2pt}
\subsc{Manual Validation}
To quantify annotation reliability, two authors independently rated a random sample of 100 detoxified comments on a 1–5 Likert scale across three dimensions (Minimal Change, Context Preservation, Communication Style), following Rahman \textit{et} al. \cite{rahmandwhp}. \revision{According to our evaluation with this 100 samples, the average score of minimal change is $4.35$, Context Preservation is $4.16$, and Communication Style is $4.48$, which indicates a higher performance of our Reframer model for our dataset}. We computed quadratic-weighted Cohen’s $\kappa$ for each dimension to respect the ordinal scale, obtaining $\kappa$ = 0.82 for Minimal Change, $\kappa$ = 0.72 for Context Preservation, and $\kappa$ = 0.77 for Communication Style. Following Landis \& Koch \cite{Landis1977}, these correspond to almost perfect agreement for Minimal Change and substantial agreement for Context Preservation and Communication Style, indicating strong consistency in our manual detoxification evaluation.

\begin{boxedtext}
\textbf{Key Finding 3:} 
\textit{\revision{ Llama 3.2 3B yielded the highest overall J-Score (84\%), effectively outperforming the other evaluated models in preserving fluency and meaning while detoxifying text.}}
\end{boxedtext}

\section{User Evaluation of ToxiShield with Integrated Browser Extension}
\label{sec:browser-etenstion}

\begin{figure}
    \centering
    \includegraphics[width=.80\linewidth]{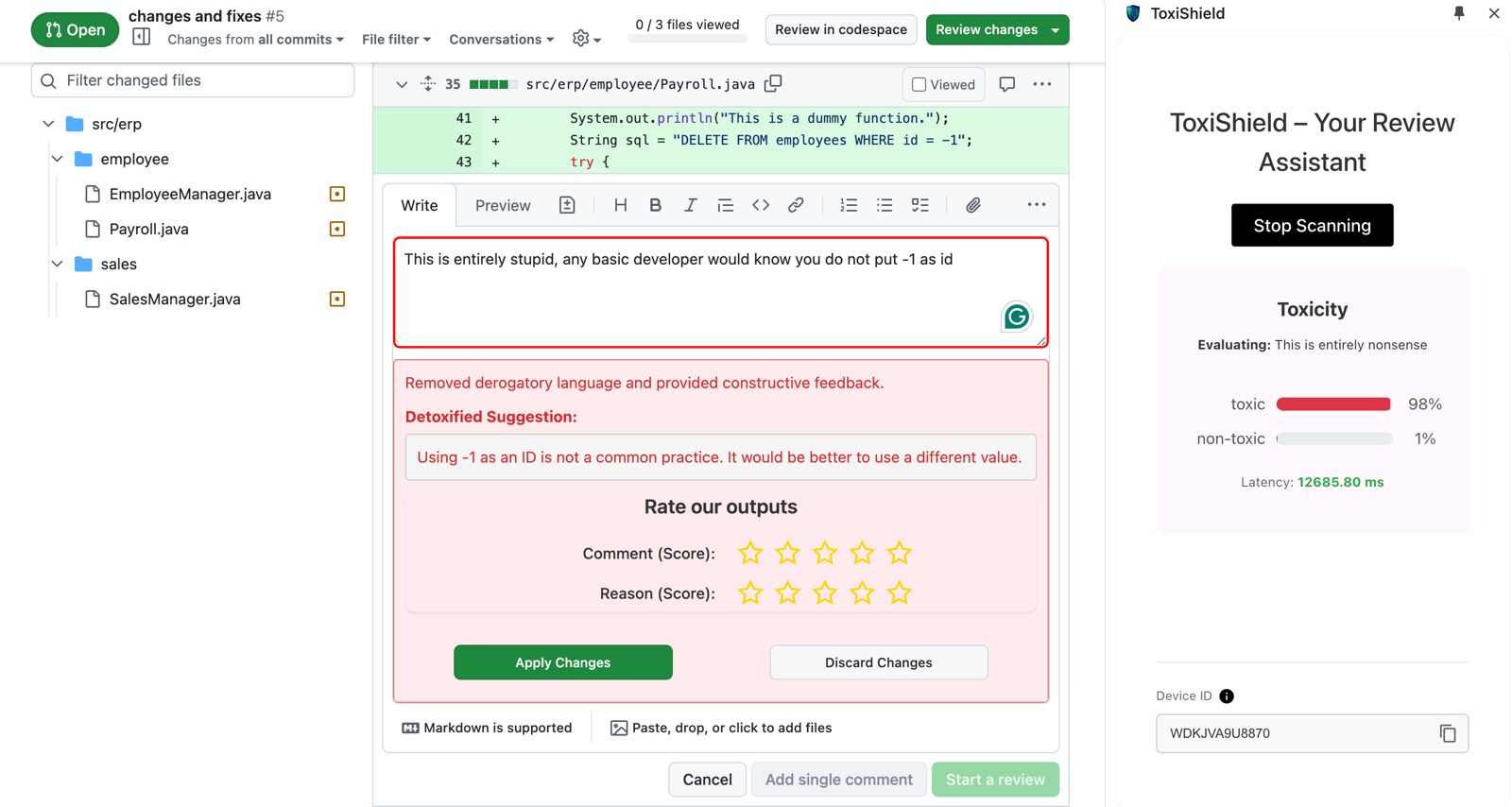}
    \caption{Browser Extension to integrate ToxiShield with GitHub-based pull request reviews}
    \label{fig:extension}
\end{figure}

We developed a browser extension (Figure~\ref{fig:extension}) designed to detect and mitigate toxic comments within GitHub pull requests in real-time. Leveraging a fully on-device architecture, the tool operates via a two-stage pipeline: first, a fine-tuned detection model monitors input as it is typed; second, upon identifying toxicity, a localized LLM generates a non-toxic alternative alongside an explanatory rationale. \revision{When a comment is classified as non-toxic, the tool takes no action and the comment proceeds to submission unchanged.} This local-first design ensures user privacy and eliminates reliance on costly external APIs. To enable this lightweight deployment, we exported our trained models to the ONNX format, ensuring cross-platform portability. Furthermore, we applied quantization techniques to significantly reduce model size and accelerate inference. These optimizations are critical for achieving the low latency required for a real-time user experience without compromising detection accuracy.

\subsection{Participant Recruitment and Survey}
\revision{Following Institutional Review Board approval, we recruited 10 professional software developers from the United States and Bangladesh to evaluate the tool's usability in a real-world setting. To ensure proficiency, participants attended a tutorial session where two authors demonstrated the installation and functionality of the browser extension. The study employed a two-week longitudinal design, during which participants integrated the tool into their daily workflows. Crucially, to strictly prioritize user privacy, we deliberately refrained from collecting telemetry or usage logs. At the conclusion of the study, participants completed a comprehensive survey consisting of both Likert-scale items and open-ended questions. Regarding usage frequency, three participants reported using the tool frequently, five occasionally, and two rarely. All survey instruments and anonymized response data are available in our replication package~\cite{ToxiShield}}.

\begin{table}[tb]

\small
\centering
\caption{Complete TAM Survey Results}
\vspace{-12pt}
\label{tab:tam-survey-results}
\resizebox{.7\textwidth}{!}{
\begin{tabular}{@{}p{9cm}cc@{}}
\toprule
\textbf{Construct / Statement} & \textbf{Mean} & \textbf{SD} \\
\midrule
\multicolumn{3}{@{}l}{\textbf{Perceived Usefulness (PU)} } \\
\midrule
The tool provides valuable features relevant to my work & 3.80 & 0.78 \\
Using this tool makes it easier to complete my tasks     & 3.40 & 0.69 \\
The tool saves me time in my daily work routine          & 3.00 & 1.05 \\
\midrule
\multicolumn{3}{@{}l}{\textbf{Perceived Ease of Use (PEOU)} } \\
\midrule
The browser extension was easy to use                    & 4.30 & 0.95 \\
I can easily navigate the tool to find different user-friendly features & 4.00 & 0.67 \\
The user interface of the tool is intuitive and user-friendly           & 4.00 & 0.67 \\
The tool provides clear and helpful error messages when needed          & 3.40 & 0.52 \\
\midrule
\multicolumn{3}{@{}l}{\textbf{Attitude Toward Use (ATU)}} \\
\midrule
Overall satisfaction (1--10 scale)                       & 7.83 & 1.83 \\
\midrule
\multicolumn{3}{@{}l}{\textbf{Behavioral Intention to Use (BI)} } \\
\midrule
I would recommend this tool to my colleagues             & 3.50 & 1.22 \\
\bottomrule
\end{tabular}}
\vspace{-12pt}
\end{table}

\subsection{Evaluation Instrument}
To rigorously evaluate developer perceptions of ToxiShield, we employed the Technology Acceptance Model (TAM)~\cite{inbook}, a widely adopted framework for assessing user adoption of new technologies. Our survey instrument incorporated nine items adapted from TAM, categorized into four distinct dimensions as detailed in Table~\ref{tab:tam-survey-results}. For the items belonging to PU, PEOU, and BI, users rated each item on a a scale of 1 to 5, where 1 means `Strongly Disagree' and 5 indicates `Strongly Agree'. The analysis of these dimensions is as follows.

\noindent \emph{i) Perceived Usefulness (PU):} This dimension measures the degree to which a user believes that using a specific system will enhance their job performance. Our results indicate that developers found ToxiShield moderately useful, particularly regarding feature relevance and task support. However, variability in time-saving metrics suggests that while the tool is helpful, further optimization is required to maximize workflow efficiency.

\noindent \emph{ii) Perceived Ease of Use (PEOU):} PEOU assesses the degree to which using the system is free of effort. ToxiShield received high ratings in this category, confirming that its design is accessible and intuitive. While the overall user experience was positive, lower mean scores related to error handling and task completion speed highlight specific usability bottlenecks that warrant attention in future iterations.

\noindent \emph{iii) Attitude Toward Use (ATU):} This metric captures the user's affective reaction to the system. We measured general satisfaction on a 10-point scale. The high aggregate score demonstrates that the majority of developers hold a positive attitude toward ToxiShield, validating its role in improving the code review experience.

\noindent \emph{iv) Behavioral Intention (BI):} BI predicts the likelihood of a user continuing to use the technology or recommending it to others. Although the reported intention was positive, the standard deviation reveals variability in user advocacy. This suggests that while the core value proposition is sound, improvements in usability and feature robustness are necessary to solidify user commitment.

\subsection{Evaluation Results}

The TAM analysis confirms that ToxiShield is generally well-received by professional developers, with particular strength in ease of use and overall satisfaction. While the tool's utility is acknowledged, enhancing task efficiency and error management mechanisms will be critical for driving broader adoption. We have a total of 8 open ended questions in our survey. Two of our authors conducted a thematic analysis of the open-ended responses following Braun and Clarke~\cite{braun2006using}, identifying four primary themes.

\noindent \textbf{Theme 1: Perceived Value and Usability.}
Participants responded positively to the tool's core concept and design. P7 described ``Blocking and refining toxic comments in real time is an innovative idea''.  P3 highlighted the value of the mission, noting that \textit{``detoxification is needed for mental health of the developers which i liked most."} Regarding usability, P2 found the tool \textit{``straightforward,"} and P6 described it as \textit{``easy to use and well build,"} validating the browser-based implementation.

\noindent \textbf{Theme 2: Workflow Integration and Latency.}
Despite the good usability, performance was a major friction point. P3 noted that \textit{``scanning is slow,"} and P7 advised \textit{``latency needs to improve a lot"} for the BERT classifier. P3 criticized the manual trigger, stating the tool should \textit{``detect pull-request or other pages on run-time"} to be truly useful. 

\noindent \textbf{Theme 3: Nuance in Toxicity Detection.}
A key finding was the tool's struggle with subtle toxicity. P5 noted that \textit{``I have used sarcastic and frustrated tones which did not come up on the toxicity radar"}. P6 provided significant insight, arguing that toxicity is often behavioral, such as \textit{``asking for unnecessary huge changes"} or \textit{``being very specific about code indentation."} P6 also emphasized that these behaviors are \textit{``hard to catch"} but can be just as harmful as harsh words.

\noindent \textbf{Theme 4: Expanding Scope Beyond Toxicity.}
Participants suggested evolving ToxiShield into a broader review assistant. P6 requested metrics for \textit{``Clarity"} and \textit{``Utility"} to measure a comment's \textit{``impact on the merge request."} P4 suggested the tool would be better if it offered \textit{``code smells"} detection, and P7 proposed a \textit{``Perhaps a chat-bot interface like Co-pilot would be more effective"} to support constructive decision-making during the writing process.

\section{Implications and Lessons Learned}
\label{sec:discussion}

\subsc{Socio Technical Implications of ToxiShield in OSS Development:} By introducing real-time, rationale-aware intervention, ToxiShield carries distinct socio-technical implications for the health of OSS communities.
First, ToxiShield fundamentally alters the economics of moderation. By automating the detection process and providing immediate, explanatory feedback, the tool significantly reduces the emotional and cognitive labor demanded of OSS maintainers, who currently shoulder the burden of manual policing. Second, the tool serves a vital pedagogical function. As ToxiShield does not merely flag content, it educates contributors by articulating why specific language is harmful. This transforms moderation from a punitive act into a learning opportunity, fostering long-term behavioral change.
Furthermore, by offering detoxified alternatives that preserve the original technical intent, ToxiShield directly supports diversity and inclusion efforts. Given that toxicity disproportionately impacts minority contributors~\cite{miller2022did}, a tool that neutralizes hostility without silencing technical critique is essential for retaining a diverse contributor base. Finally, recognizing that Codes of Conduct (CoC) are often insufficient on their own~\cite{bharadwajshifting}, ToxiShield acts as an operational enforcement mechanism, bridging the gap between community ideals and daily practice. By embedding these values directly into the communication pipeline, ToxiShield can help cultivate a more resilient and inclusive OSS ecosystem.

\vspace{2pt}
\subsc{Effectiveness of LLMs in Code Review Detoxification: } We evaluated seven fine-tuned LLMs on our detoxification dataset using four complementary metrics: style accuracy, fluency, content preservation, and the aggregated J-score (Table~\ref{tab:student_metrics}). All models reduced toxicity by more than 87\% as measured by ToxiCR, confirming that LLMs are effective detoxification tools. However, their ability to preserve semantic content varied considerably. These findings indicate that while most LLMs perform well in toxicity reduction and fluency (all exceeding 94\% in fluency), semantic preservation remains a challenge, with even the best models retaining only about two-thirds of the original meaning. Model selection thus critically affects practical utility: lightweight, open-source models like Llama 3.2 3B can offer a strong balance of effectiveness, efficiency, and deployability for real-time developer tooling.

\vspace{2pt}
\subsc{Open-Source LLMs vs Proprietary Models:} Our evaluation shows fine-tuned open-source models can match or surpass proprietary systems. While GPT-4o mini achieved the highest style accuracy, Llama-3.2 3B secured the best overall score by better preserving content. This demonstrates that small, open-source models are ideal for real-time deployment: they are cost-effective, transparent, and locally hosted. By eliminating reliance on external APIs, these lightweight checkpoints bypass common privacy and compliance hurdles, making ToxiShield practical for widespread enterprise adoption. Future production deployments will incorporate telemetry to provide longitudinal insights into adoption and ToxiShield's long-term impact on developer communication norms.

\section{Threats to Validity}
\label{sec:threats}
\textbf{Internal Validity} There is a risk that the models, particularly those fine-tuned using LoRA, may overfit to the specific dataset used in this study, potentially limiting their generalizability to other datasets or domains. Although cross-validation was employed to mitigate this risk, it cannot be entirely eliminated.
Moreover, our multi-class evaluation was conducted on a small-scale dataset of 1,200 samples, including the only publicly available corpus offering fine-grained toxicity labels compatible with our 11-class schema~\cite{sarker2025landscape}.
Future studies should evaluate larger, more diverse corpora to mitigate the limited statistical power and sampling bias of our dataset.

\vspace{2pt}
\subsc{External Validity}
First, our dataset's focus on code reviews, with its specific jargon and code snippets, may limit the applicability of our findings to other domains. However, since the data is from GitHub, our results should remain valid within the broader software engineering community. Second, our models were optimized for a single task (toxicity classification) and may not transfer effectively to other NLP applications or software contexts.
Finally, a potential threat to the generalizability of our findings is the sample size of our user study ($N=10$). Our recruitment was constrained by strict IRB limitations on geographic eligibility and by a high barrier to entry (a two-week commitment and installation of a browser plugin). However, we mitigated this threat by ensuring deep engagement; all participants provided extensive feedback over the two-week period. Furthermore, within the context of usability evaluation, prior research posits that 5 participants are sufficient to identify 80\% of usability issues~\cite{virzi1992refining}, and 10 participants can uncover up to 92\% of issues~\cite{faulkner2003beyond}. Thus, while our sample size limits broad demographic generalization, we believe it is sufficient to capture the primary usability and workflow challenges of the tool.

\vspace{2pt}
\subsc{Construct Validity}
First, our definition of toxicity~\cite{miller2022did, sarker2023automated} may not encompass subtle forms of toxic behavior, such as passive-aggressive comments. Second, our use of prompt tuning could lead to overly dependent results on specific prompts, thereby affecting reproducibility. To mitigate the related risk of LLM non-determinism, we fixed generation parameters (e.g., temperature, top-p) and conducted a structured error analysis, which increased the consistency and trustworthiness of the detoxification process.

\vspace{2pt}
\subsc{Ecological Validity} While our real-time toxicity tool has proven effective in controlled tests, its practical success depends on user adoption and seamless workflow integration. As the first real-time detoxification tool built for the SE domain, it’s a key step toward healthier developer communication. We will use feedback from real-world deployment to address challenges like alert fatigue, continuously improving the tool's adoption and impact.

\section{Related Work}
\label{sec:background}

\noindent \textbf{Antisocial Behavior in Open Source Software Development Environments}:
Antisocial behavior in the SE domain is broadly defined as actions that harm community participation~\cite{Qiu2022,miller2022did,sarker2020benchmark,sarker2025landscape}. Although OSS environments exhibit lower toxicity than general social media (i.e., Reddit or Twitter), their professional and technical dynamics make these interactions qualitatively distinct~\cite{miller2022did}. Toxicity often manifests subtly through personal attacks (`destructive criticism'~\cite{gunawardena2022destructive}), aggressive rejections (`pushback'~\cite{egelman2020predicting}), or dismissive tones (`incivility'~\cite{ferreira2021shut}). Additionally, a recent large-scale study~\cite{sarker2025landscape} revealed that toxicity is unevenly distributed, peaking in gaming projects and strongly correlates with technical factors such as code churn and review intervals. Building on this extensive foundation, our study moves beyond characterization to develop adaptive, real-time strategies to mitigate toxic interactions in OSS environments.

\begin{table}
    \caption{Feature comparison highlighting ToxiShield's comprehensive capabilities. Checkmarks (\cmark) indicate support, while crosses (\xmark) indicate unsupported features. MC-> Multi Class, DT-> Detoxification}
    %\resizebox{\textwidth}{!}{
\centering
\small

\newcolumntype{M}[1]{>{\centering\arraybackslash}m{#1}}

\begin{tabular}{|p{3.2cm}|M{1.5cm}|M{1.4cm}|M{1.4cm}|M{2.1cm}|M{2cm}|}
\hline
\textbf{Tool} & \textbf{Binary} & \textbf{MC} & \textbf{DT} & \revision{\textbf{Explanation}} & \revision{\textbf{Real Time}} \\
\hline
Raman \textit{et} al.~\cite{raman2020stress}        & \cmark & \xmark & \xmark & \xmark & \xmark \\
\hline
Egelman \textit{et} al.      & \xmark & \xmark & \xmark & \xmark & \xmark \\
\hline
ToxiCR~\cite{sarker2023automated} & \cmark & \xmark & \xmark & \xmark & \xmark \\
\hline
ToxiSpanSE~\cite{sarker2023toxispanse} & \cmark & \xmark & \xmark & \xmark & \xmark \\
\hline
Ferreira \textit{et} al.~\cite{ferreira2024incivility}     & \xmark & \xmark & \xmark & \xmark & \xmark \\
\hline
Rahman \textit{et} al.~\cite{rahmandwhp}       & \cmark & \xmark & \cmark & \xmark & \xmark \\
\hline
Mishra \&  Chatterjee~\cite{chatgpt-github-toxicity}       & \xmark & \xmark & \xmark & \xmark & \xmark \\
\hline
\textbf{ToxiShield} & \cmark & \cmark & \cmark & \cmark & \cmark \\
\hline
\end{tabular}
%}
\vspace{-15pt}
    \label{tab:tool_comparison}
    \end{table}

\vspace{2pt}
\noindent \textbf{Automated Detection of Toxicity in Software Engineering}:
General-purpose Natural Language Processing (NLP) tools often fail to accurately detect toxicity within Software Engineering (SE) contexts due to their inability to parse technical jargon and developer-specific communication styles~\cite{novielli2021assessment,sarker2020benchmark}. Recognizing this gap, Raman~\textit{et al.}~\cite{raman2020stress} pioneered the field by developing the first toxicity detector specifically tailored for OSS communities. However, subsequent evaluations revealed significant limitations in this early model's generalizability and accuracy~\cite{miller2022did, Qiu2022, sarker2020benchmark}. Addressing these shortcomings, later studies sought to refine detection capabilities. Qiu~\textit{et al.}~\cite{Qiu2022} designed a classifier that integrated metrics for `pushback'~\cite{egelman2020predicting} and `toxicity'~\cite{raman2020stress}, while Sarker~\textit{et al.}~\cite{sarker2023automated} established a new benchmark with ToxiCR, a tool trained on 19,651 manually labeled code review comments that achieved an $88.9\%$ F1-score. Most recently, Rahman~\textit{et al.}~\cite{rahmandwhp} shifted the paradigm toward generative AI, utilizing both vanilla and fine-tuned Large Language Models (LLMs) to not only identify uncivil comments but also propose civil alternatives. Our work builds upon these foundational efforts, advancing the state of the art by integrating accurate detection with real-time, explanatory mitigation strategies.

\vspace{2pt}
\noindent \textbf{Repercussions of toxicity on OSS projects}: The consequences of toxic interactions in OSS development extend far beyond momentary unpleasantness, creating a cascade of negative effects that damage both individuals and communities. Research demonstrates that toxicity fundamentally undermines developers' mental health~\cite{carillo2016towards}, serving as a primary driver of significant stress and burnout~\cite{raman2020stress}. These individual struggles inevitably degrade the collective workflow; phenomena such as unjustified pushback and destructive criticism have been shown to diminish productivity and severely strain interpersonal relationships~\cite{egelman2020predicting, murphy2022pushback}. Most critically, toxicity disproportionately harms minority developers, acting as a systemic barrier that obstructs Diversity, Equity, and Inclusion efforts~\cite{gunawardena2022destructive}. Building upon this evidence of harm, our study proposes actionable strategies to mitigate these adverse effects and foster healthier OSS communities.

\vspace{2pt}
\noindent\textbf{Novelty:}
ToxiShield distinguishes itself from existing toxicity detection approaches through three advancements, as summarized in Table~\ref{tab:tool_comparison}. First, it moves beyond binary classification by employing multi-class analysis to identify toxicity subcategories. Second, it provides explanatory feedback detailing \textit{why} a comment is toxic to foster behavioral change. Finally, ToxiShield represents the first empirically evaluated, real-time toxicity mitigation tool integrated into developer workflows.

\section{Conclusion}
\label{sec:conclusion}
Our study introduces a robust approach for detecting and mitigating toxicity in code review environments, fostering healthier communication in software development. We present ToxiShield, a domain-specific dataset, and fine-tune state-of-the-art language models using LoRA and prompt-tuning techniques to address the nuances of toxicity in code review discourse. Our pipeline enables real-time detection, multiclass classification, and detoxification, with ToxiShield offering cost-efficient integration into existing platforms. Experimental results demonstrate the tool’s effectiveness, though challenges like overfitting and limited generalizability exist. This work lays the groundwork for research in automated toxicity detection and inclusive developer collaboration.

\section*{ACKNOWLEDGEMENT}
\vspace{-4pt}
\label{acknowledge}
This research is partially supported by the US National Science Foundation under Grant No. 2340389, and a start-up grant from the University of Nebraska at Omaha, College of Information Science and Technology. The findings of this research do not necessarily reflect the views of the National Science Foundation. We are also grateful to the Applied Machine Learning Lab in the Department of CSE at Bangladesh University of Engineering and Technology for logistical support. 

\section*{DATA AVAILABILITY}
\label{Data_availability}
\vspace{-4pt}
The dataset, source code, and experimental results of ToxiShield are available here ~\cite{ToxiShield} and DOI: \url{https://doi.org/10.5281/zenodo.19490337}. 

\bibliographystyle{ACM-Reference-Format}  

\bibliography{bibliography}

\end{document}